\documentclass[12pt]{article}

\advance\hoffset by -1.1truecm
\advance\voffset by -1.0truecm

\def\be{\begin{equation}}
\def\ee{\end{equation}}
\def\ba{\begin{eqnarray}}
\def\ea{\end{eqnarray}}

\newcommand{\x}{{\bf x}}
\newcommand{\p}{{\bf p}}
\newcommand{\sn}{\smallskip\newline}
\newcommand{\mn}{\medskip\newline}

\begin{document}
\title{Mode Generating Mechanism in Inflation with Cutoff}
\author{Achim Kempf\\
Institute for Fundamental Theory\\ Departments
 of Physics and Mathematics\\
University of Florida, Gainesville, FL 32611, USA\\
{\small Email:  kempf@phys.ufl.edu}}

\date{}

\maketitle

\vskip-7.5truecm

\hskip11.5truecm
{\tt UFIFT-HEP-00-25}

\hskip11.5truecm {\tt astro-ph/0009209} \vskip7.1truecm

\begin{abstract}
In most inflationary models, space-time inflated to the extent that
modes of cosmological size originated as modes of wavelengths 
at least several orders of magnitude smaller than the Planck length.
Recent studies confirmed that, therefore, 
inflationary predictions for the cosmic microwave background 
perturbations are 
generally sensitive to what is assumed about the
Planck scale. Here, we 
propose a framework for field theories on  
curved backgrounds with a plausible type of ultraviolet cutoff.
We find an explicit mechanism by which during cosmic expansion
new (comoving) modes are generated continuously. Our results 
allow the numerical calculation of a prediction for
the CMB perturbation spectrum.
\end{abstract}
Several problems 
of standard big bang cosmology, such as
the horizon and the flatness problems, can be 
explained under the assumption that   
the very early universe underwent a period
of extremely rapid inflation, driven 
by the potential of some assumed inflaton field.
In particular, the inflationary scenario
is also able to explain the observed perturbations in the 
cosmic microwave background (CMB), 
namely as  originating ultimately from quantum 
fluctuations of the inflaton field. Indeed,
the inflationarily predicted gaussianity 
and near scale invariance of the perturbations'
spectrum closely matches the current experimental 
evidence, see e.g. \cite{peacock}.
\sn
However, it has also been pointed out that in typical inflationary 
models, such as simple models of chaotic inflation, space-time 
inflated to the extent that modes which are now of cosmological size
originated as modes with wavelengths that were at least several orders of
magnitude smaller than the Planck length. 
Until recently, reason to believe that the inflationary 
prediction
of the CMB spectrum might be insensitive to structure at the Planck 
scale was provided by the analogy with black hole radiation, which
suffers from a similar
transplanckian problem: 
any asymptotic 
Hawking photon with a medium range frequency 
should have had a far transplanckian proper frequency
close to the event horizon, even at distances
{} from the horizon which are 
farther than a Planck length.
This problem has been investigated in detail, see e.g.  
\cite{bhrad}. The current 
consensus appears to be that the derivation of
Hawking radiation is indeed essentially
robust to changes in the assumed Planck scale structure. 
Likely, the reason for this insensitivity is that there are 
basic thermodynamical arguments for the properties of
black hole radiation.
\sn
In the case of inflation, however,
there does not seem to exist
any basic thermodynamic reason why the derivation of the
particle production should be insensitive 
to what is assumed about the Planck scale.
Indeed, recent studies, see \cite{brand,niemeyer},
independently found that the inflationary CMB
prediction \it is \rm in general sensitive.
Those studies calculated the effects of ad hoc ultraviolet
modified dispersion relations
similar to those which had been shown not to affect black hole 
radiation.
\sn
Interestingly, this means that if the inflationary scenario is true, then
the currently rapidly improving experimental evidence about the CMB could
provide a window to at least some aspects of particle physics at energies
as high as the Planck scale.
Therefore, our aim here is to formulate 
quantum field theory 
on curved backgrounds with some plausible type of ultraviolet cutoff. 
The aim is to obtain a framework in which all the 
usual entities of interest such as
propagators, correlators, vacuum fluctuations and eventually the
CMB perturbation spectrum can 
be calculated explicitly and compared to experiment.
\sn
We will need good arguments for
choosing a particular type of cutoff. One obvious option
would be to choose the
lattice cutoff. However, apart from breaking
translation and rotation invariance, the lattice
cutoff faces further problems on expanding space-times: 
As has been pointed out earlier, see e.g. \cite{niemeyer}, if
space-time were a discrete lattice with a spacing of say 
one Planck length then
it is not clear how during the expansion 
new discrete lattice points could 
be created continuously.  
Indeed, whatever form of natural ultraviolet cutoff we assume,  
we will need to address the following questions:
How does the expansion continuously generate new modes? 
What is the
initial vacuum of these modes? And, once this is clarified, 
which CMB perturbation spectrum is predicted?
\mn
Our starting point for choosing a type of cutoff 
is an observation made in \cite{ak-erice}:
The short-distance structure of space-time can only be  one
of very few types - if we make a certain assumption. The assumption 
is that the fundamental 
theory of quantum gravity will possess for each 
space-time coordinate 
a linear operator $X^i$ 
whose formal expectation values $\langle X^i\rangle$
are real.  
The $X^i$ may or may not commute. 
One can prove on functional analytic grounds that
the short-distance structure of any such coordinate, 
considered separately, can only be continuous, discrete, or ``unsharp" in 
one of two ways. All other cases are mixtures of these.
\sn 
The mathematical origin of the two ``unsharp" cases lies in the fact that
such an operator
is not necessarily self-adjoint and may be merely what is
called a symmetric operator.
If $X^i$ is self-adjoint, then it is diagonalizable and its spectrum 
can only be discrete (i.e., point) 
or continuous. Correspondingly, the short-distance 
structure of the coordinate which is described by $X^i$ is therefore also
discrete or continuous (or a mixture, including, e.g., fractals).
However, if $X^i$ is merely symmetric, then, crucially,
it is not diagonalizable. 
Such coordinates are unsharp because
a merely symmetric operator $X^i$ need not possess
eigenvectors, and the eigenvectors of 
its adjoint $(X^{i})^*$
need not be orthogonal. (Technically, the two unsharp cases are 
distinguished
by the so-called deficiency indices of $X^i$
being either equal or unequal, see \cite{ak-erice}.)
\sn
Given the generality of the argument, namely the fact that the 
sharpness or
 unsharpness of indeed {\it any}  real entity described by a linear operator
falls into this classification, it is not surprising that  
such unsharp short-distance structures do occur ubiquitously. For example, 
the aperture induced unsharpness of optical images is of this kind, as
 is, e.g., the unsharpness in the
time resolution of bandlimited electronic signals, see \cite{ak-shannon}.
\sn
In fact, also a number of studies in quantum gravity and string
theory point towards one
of these unsharp cases: They point towards the
 case of coordinates $X$ whose formal
uncertainty $\Delta X(\phi)=\langle \phi \vert 
(X - \langle \phi\vert X\vert \phi\rangle)^2\vert\phi\rangle^{1/2}$   
possesses a finite lower bound at a Planck or string scale,
$\Delta X(\phi)\ge \Delta X_{min} = l_{Pl}$, where $\phi$ is any unit vector 
 on which the operator $X$ can act. (Technically,
 this is a case of equal deficiency indices.)
\sn
In a first-principles quantum gravity theory such
 as string theory this behavior
may of course arise from a
complicated dynamics where space-time is a derived concept.
Nevertheless, it has been argued, 
see e.g. \cite{grossetal}, that, effectively,
this short-distance structure can be modeled as
arising from quantum gravity correction terms
to the uncertainty relation: 
\be
\Delta x \Delta p \ge \frac{1}{2} (1+ \beta (\Delta p)^2 + ...) 
\label{ucr}
\ee
The positive constant $\beta$ implies
a constant positive lower bound
$\Delta x_{min} = \sqrt{\beta}$. Usually, $\beta$
is assumed such that the cutoff 
$\Delta x_{min}$ is at a Planck or string scale.  
As first discussed in \cite{ak-jmp-ucr}, the type of
uncertainty relation of Eq.\ref{ucr} can then be viewed
as arising from corrections to the canonical 
commutation relations:
\be
[\x,\p]=i (1+\beta \p^2 ...)
\label{nccr}
\ee
This commutation relation can be 
represented, e.g., with the operators $\x$ and $\p$ acting 
on fields over some auxiliary variable $\rho$ 
as: $\x \phi(\rho)
=i\partial_\rho \phi(\rho),~~\p \phi(\rho) = 
\tan(\rho \sqrt{\beta})/\sqrt{\beta}$ on 
the space of fields
 $\phi(\rho)$ over the interval $[-\rho_{max},
\rho_{max}]$, where
$\rho_{max}=\pi/(2\sqrt{\beta})$, 
with scalar product $\langle\phi_1\vert\phi_2\rangle=
\int_{-\rho_{max}}^{\rho_{max}}d\rho 
\phi_1^*(\rho)\phi_2(\rho)$. The requirement of
symmetry of $\x$ (i.e.,  that
all formal expectation values are real) yields 
the boundary conditions $\phi(\pm\rho_{max})=0$.
As expected, $\x$ is not self-adjoint:
what would be its (now normalizable) 
eigenvectors ``$\vert x)$", namely 
the plane waves in $\rho$-space, are not in its domain and are
not orthogonal on the
interval $[-\rho_{max},\rho_{max}]$.
In the ``position representation", the fields $\phi(x) =(x,\phi)=
\pi^{-1/2}\beta^{1/4}\int d\rho ~e^{i x \rho}$ exhibit a 
finite minimum wavelength.
Modes whose wavelengths are close to the minimum wavelength 
are energetically exceedingly expensive ($p$ diverges).
\sn
It would be a drawback of our approach if it
covered only this perfectly translation
invariant case in which Fourier theory applies. 
Let us therefore also mention, for completeness,
that in this class of unsharp short-distance structures
the lower bound on the position 
uncertainty, $\Delta X_{min}$,
is in general some function of the position $\langle X\rangle$ 
around which one tries to
localize the particle:  $\Delta X_{min}=\Delta 
X_{min}(\langle X\rangle)$. In this  
case, the situation is no longer describable as
 an overall wavelength cutoff. 
But it has been shown in \cite{ak-shannon}, that in all cases where
$ \Delta X_{min}(x)>0$,
fields over such unsharp coordinates can be
 described as being ultraviolet cutoff 
in the sense that they
contain only a finite density of
 degrees of freedom. Namely, such fields
can be reconstructed at all points if known only
 on a set of discrete points - if these
points are sufficiently tightly spaced. The 
minimum spacing varies over space and 
is small where the minimum position uncertainty
 $\Delta X_{min}(x)$ is small, and vice versa.
Thus, fields over such unsharp coordinates, while 
being continuous, always possess regularity
 properties similar to fields over lattices. 
\mn
We now begin by recalling that the calculation
 of inflationary scalar density perturbations 
effectively reduces in the simplest case
to the study of a minimally coupled massless
 real scalar field on a fixed curved background
space-time such as, e.g., de Sitter space.
For simplicity, we will assume the case of spatial
 flatness. It is usually most
convenient to choose comoving coordinates 
${y}$ and the conformal time 
coordinate $\eta$. The metric then reads 
$g= $diag$(a(\eta)^2,-a(\eta)^2,
-a(\eta)^2,-a(\eta)^2)$ with $a(\eta)$ being 
the scale factor, and the action takes the form:
\be
S=\int d\eta~d^3y~\frac{a(\eta)^2}{2}\left((\partial_\eta \phi)^2
-\sum_{i=1}^3(\partial_{y^i}\phi)^2\right)
\label{stac}
\ee
Since our aim is to introduce a short-distance
 cutoff in proper distances (while leaving
the time coordinate as is), we
transform the action into proper space 
coordinates $x^i$, to obtain 
\be
S=\int d\eta ~d^3x ~\frac{1}{2a}\left\{\left[\left(\partial_\eta + 
\frac{a'}{a}~\sum_{i=1}^3\partial_{x^i}x^i
 - \frac{3a'}{a}\right)\phi\right]^2 - a^2
\sum_{i=1}^3\left(
\partial_{x^i}\phi\right)^2\right\}
\ee
where ${}^\prime$ stands for $\partial_\eta$. By defining
\begin{eqnarray}
(\phi_1,\phi_2) & := & \int d^3x ~ \phi^*_1(x)\phi_2(x)\\
\x^i\phi(x) & := & x^i \phi(x)\\
 \p^i\phi(x) & := & -i\partial_{x^i}\phi(x)
\end{eqnarray}
we can write the action in the form
\be
S=\int d\eta~\frac{1}{2a}\left\{\left(\phi,A^\dagger(\eta)
A(\eta)\phi\right)-a^2\left(\phi,\p^2\phi\right)
\right\}
\label{ab}
\ee
where the operator $A(\eta)$ is defined as: 
\be
A= 
\left(\partial_\eta + i 
\frac{a'}{a}~\sum_{i=1}^3\p^i\x^i - \frac{3a'}{a}\right)
\label{abac}
\ee
In Eq.\ref{ab}, the fields are time dependent 
abstract vectors in a Hilbert space representation 
of the commutation  relations
\be
[\x^i,\p^j]=i \delta^{ij} .
\label{qmccr}
\ee
This only means that, as in quantum mechanics, 
also in quantum field theory the commutation 
relations of Eq.\ref{qmccr} are setting
the stage, albeit without the simple quantum 
mechanical interpretation of the $\x^i$ and $\p^i$ as observables. For example,
to express the action in momentum space is to choose the spectral
representation of the $\p^i$.
\sn
Our ansatz now is to the keep the scalar field
 action exactly as given in
Eqs.\ref{ab},\ref{abac}, but to modify the underlying three-dimensional 
position-momentum commutation relations Eqs.\ref{qmccr} 
for large momenta, such 
as to introduce the type of cutoff  which we
 discussed above. While we will 
break Lorentz invariance by introducing the cutoff, we will 
maintain translation and rotation invariance through
 the ansatz 
\be
[\x^i,\p^j]=i\left(f({\p}^2)\delta^{ij} + g({\p}^2) \p^i\p^j\right)
\label{nccrs}
\ee
and by requiring that $[\x^i,\x^j]=0=[\p^i,\p^j]$, for
all $i,j=1,2,3$. As was first shown in \cite{ak-osc}, 
the Jacobi identities then relate the functions $f$ and $g$
as follows:
\be
g = \frac{2 f \partial_{p^2}f}{
f-2p^2\partial_{p^2}f}
\label{fg}
\ee
The behavior of the functions $f$ and $g$ for $p^2$ small compared
to the Planck momentum  
is required to be $f\rightarrow 1$ and $g \rightarrow 0$. 
The functions $f$ and $g$ are then unique to first order in $\beta$, namely:
$f=1+\beta{p}^2+O(\beta^2)$ and $g= 2 \beta+O(\beta^2)$. 
We note that corresponding to the ambiguity in 
choosing $f$ there is also an ordering ambiguity
of the $\x^i$ and $\p^j$ in the action of Eqs.\ref{ab},\ref{abac}.
Eq.\ref{fg} shows that $g$ may develop singularities. We avoid this by
choosing, e.g.,  $g = 2 \beta$. This then yields from Eq.\ref{fg} that
$f(p^2)=2\beta {p}^2/(\sqrt{1+4\beta {p}^2}-1)$. 
A convenient Hilbert space representation 
of the new commutation relations
is on fields $\phi(\rho)$ over auxiliary variables $\rho^i$
\be
\x^i\phi(\rho) =  i\partial_{\rho^i}\phi(\rho)
\ee
\be
\p^i\phi(\rho) = \frac{\rho^i}{1-\beta{\rho}^2}\phi(\rho)
\ee
with scalar product: 
\be
(\phi_1,\phi_2) = \int_{{\rho}^2<\beta^{-1}}
d^3\rho~\phi_1^*(\rho)\phi_2(\rho)
\ee
The symmetry of the $\x^i$ requires the boundary condition
$\phi(\rho^2=1/\beta)=0$. The $\x^i$ are not self-adjoint: 
Their would-be eigenvectors ``$\vert x)$", i.e., 
the now normalizable
plane waves in $\rho$-space, are not in their domain and
are not orthogonal. 
The finiteness of $\rho$-space, 
$\rho^2_{max}=\beta^{-1}$, implies  
a finite minimum wavelength 
$\lambda_{min}=2\pi\sqrt{\beta}$
for the fields $\phi(x)=(x,\phi)= \sqrt{3/(4\pi)} \beta^{3/4}
\int d^3\rho~\phi(\rho) e^{i x \rho}$
over position space, 
and a corresponding finite minimum 
position uncertainty.
\sn
The action, as given in its abstract form, 
Eqs.\ref{ab},\ref{abac}, with the new  
commutation relations Eqs.\ref{nccrs}
 underlying, can now be written in the $\rho$-representation:
\be
S=\int d\eta\int_{{\rho}^2<\beta^{-1}}d^3\rho~\frac{1}{2a}\left\{
\left\vert \left(\partial_\eta -\frac{a'}{a}\frac{
\rho^i}{1-\beta {\rho}^2}\partial_{\rho^i}
-\frac{3a'}{a}\right)\phi
\right\vert^2 -\frac{a^2 {\rho}^2\left\vert 
\phi\right\vert^2}{(1-\beta {\rho}^2)^2} \right\}
\ee
The presence of  $\rho$ derivatives means that the 
$\rho$ modes are coupled. Fortunately,
it is still possible to find new variables $(\tilde{\eta},{\tilde{k}})$,
namely
$\tilde{\eta}=\eta,~ \tilde{k}^i=a\rho^i \exp(-\beta {\rho}^2/2)$,
in which the $\tilde{k}$ modes decouple. As is readily verified:
$\partial_{\eta} - \frac{a'}{a}\frac{\rho^i}{1
-\beta{\rho}^2}\partial_{\rho^i} =\partial_{\tilde{\eta}}$.
We will use the common index notation
 $\phi_{\tilde{k}}$ for those decoupling modes.  
The realness of the
field $\phi(x)$ then 
translates through $\phi(\rho)^*=\phi(-\rho)$ 
into $\phi^*_{\tilde{k}}=\phi_{-\tilde{k}}$.
We observe that the $\tilde{k}$ modes
coincide 
with the usual comoving modes that are obtained by scaling,
$k^i=a p^i$, only on large scales, i.e., only for small $\rho^2$, i.e., only for small
 momentum eigenvalues $p^2$.
Conversely, this means that
the comoving $k$ modes only decouple at large proper distance scales, 
but do couple at small scales.
\sn
The action now reads 
\be
S = \int d\eta \int_{\tilde{k}^2<a^2/e\beta}
 d^3\tilde{k}~ 
{\cal L}
\label{nac1}
\ee 
with
\be
{\cal L} = \frac{1}{2}\nu
\left\{\left\vert\left(\partial_{\eta} - 
3\frac{a'}{a}\right)\phi_{\tilde{k}}(\eta)
\right\vert^2
- \mu
\vert\phi_{\tilde{k}}(\eta)
\vert^2\right\}
\label{nac2}
\ee
where we defined 
\begin{eqnarray}
\mu(\eta,\tilde{k}) & := & 
-\frac{a^2 \mbox{plog}(-\beta \tilde{k}^2/a^2)}{
\beta\left(1+\mbox{plog}\left(-\beta \tilde{k}^2/a^2\right)\right)^2}
\\
\nu(\eta,\tilde{k}) & := & 
\frac{e^{-\frac{3}{2}\mbox{plog}\left(-\beta \tilde{k}^2/a^2\right)}}{
a^4 \left(1+\mbox{plog}\left(-\beta \tilde{k}^2/a^2\right)\right)}
\end{eqnarray}
and where 
the function plog, the ``product log", being the inverse
 of the function $x\mapsto x e^x$, 
allowed us to express $\rho^2$ in terms of $\eta$ and $\tilde{k}^2$ through:
\be
\rho^2=-\beta^{-1}\mbox{plog}(-\beta\tilde{k}^2/a^2).
\ee
Before we proceed, let us remark that this action
reduces for $\beta\rightarrow 0$ to the standard action: 
\be
S_{{}_{\beta=0}}=\int d\eta d^3k~\frac{1}{2 
a^4}\left\{\left\vert\left(\partial_\eta-3\frac{a'}{a}\right)\phi
\right\vert^2-k^2\vert \phi\vert^2\right\}
\label{coac}
\ee
To see that this is the standard action, recall that 
Fourier transforming and scaling commute
only up to a scaling factor. We Fourier transformed the proper
positions and then scaled. 
Thus, our $\phi$  differs by a factor of $a^3$ 
{} from the $\phi$ obtained as usual
by first scaling and then Fourier transforming.
\sn
The equation of motion for the action
 of Eqs.\ref{nac1},\ref{nac2} is:
\be
\phi_{\tilde{k}}^{\prime\prime} + \frac{\nu^\prime}{\nu} 
\phi_{\tilde{k}}^\prime +\left(
\mu
 - 3 \left(\frac{a^\prime}{a}\right)^\prime 
-9\left(\frac{a^\prime}{a}\right)^2
-\frac{3 a^\prime \nu^\prime}{a \nu}
\right)\phi_{\tilde{k}} = 0
\label{eom}
\ee
We recognize the damped harmonic oscillator
 form, with its friction term and with its
variable mass term in which the
 contributions from the momentum and the expansion 
compete. The field $\pi_{\tilde{k}}(\eta)$, 
canonically conjugate to $\phi_{\tilde{k}}(\eta)$, reads: 
\be
\pi_{\tilde{k}}(\eta) = \nu \phi^\prime_{-\tilde{k}}(\eta)
 - 3 \nu \frac{a^\prime}{a}\phi_{-\tilde{k}}(\eta)
\ee
(recall that the canonical conjugate of the 
Fourier transform is the
complex or hermitian conjugate of the Fourier transform of the 
canonical conjugate). 
We can now use that $\phi^\prime_{\tilde{k}}(\eta)
= \nu^{-1} \pi_{-\tilde{k}}(\eta) + 
3 \frac{a^\prime}{a}\phi_{\tilde{k}}(\eta)$ 
to express the Hamiltonian
\be
H= \int_{\tilde{k}^2<a^2/e\beta} d^3\tilde{k}~ 
\left(\pi_{\tilde{k}}(\eta)
 \phi^\prime_{\tilde{k}}(\eta) -{\cal 
L}\right)
\label{ham}
\ee
explicitly in terms of $\phi$ and $\pi$. We  
quantize by imposing 
the commutation relation
$[\hat{\phi}_{\tilde{k}}(\eta),\hat{\pi}_{\tilde{r}}(\eta)] 
= i \delta^3(\tilde{k}-\tilde{r})$.
(A path integral formulation of field theories over
unsharp coordinates has been developed
in \cite{path}.)
The Heisenberg
 equations $\hat{\phi}_{\tilde{k}}^\prime=i[\hat{H},
\hat{\phi}_{\tilde{k}}] $ 
and $\hat{\pi}_{\tilde{k}}^\prime=i
[\hat{H},\hat{\pi}_{\tilde{k}}]$ then
yield Eq.\ref{eom} as an equation for operator-valued 
fields. 
\sn
We can now answer the first question which we raised
in the beginning, namely by which mechanism 
new modes are generated:
Automatically,
the quantum field $\hat{\phi}(\eta,x)$ and   
the quantum Hamiltonian $\hat{H}$ of Eq.\ref{ham}
contain the $\tilde{k}$ mode, i.e., the fields
$\hat{\phi}_{\tilde{k}}(\eta)$ and 
$\hat{\pi}_{\tilde{k}}(\eta)$, only
after the $\tilde{k}$ mode's `creation' time, $\eta_c(\tilde{k})$, 
which is when 
$a(\eta)$ has grown
enough so that $\tilde{k}^2<a^2/e\beta$, 
i.e.,  when
the proper wavelength of the $\tilde{k}$ mode becomes 
larger than $\lambda_{min}$.
The action of $\hat{H}$ on the Hilbert subspace of 
modes with $\tilde{k}^2>a^2/(e\beta)$ is zero, i.e.,  
the time evolution 
operator leaves the respective Hilbert 
subspace invariant.
\sn
Technically, the kernel of the Hamiltonian (its eigenspace 
to eigenvalue $0$) is 
infinite dimensional and shrinks during cosmic expansion.
Conversely, during a cosmic contraction
 the kernel enlarges. It should be interesting
to calculate the correspondingly changing
 zero-point energy of the Hamiltonian as it
picks up or loses modes. In this scenario, we live quite
literally in a universe which resembles 
``Hilbert's hotel" (which can 
welcome guests even if full,  because it has 
an infinite number of rooms).
\sn
We can solve for the 
dynamics of the quantized field $\hat{\phi}$
as usual, by using a complex classical solution $\phi$ 
to write the $\tilde{k}$ mode of the 
quantum field $\hat{\phi}$ as:
\be
\hat{\phi}_{\tilde{k}}(\eta) 
= \left(a_{\tilde k} \phi_{\tilde{k}}(\eta)+a^\dagger_{-\tilde{k}}
\phi_{-\tilde{k}}^*(\eta)\right)
\ee 
The time independence of the $a_{\tilde{k}}$ and 
$a_{\tilde{k}}^\dagger$   
guarantees that $\hat{\phi}$ solves
the equation of motion. Imposing $[a_{\tilde{k}},
a^\dagger_{\tilde{r}}]=
\delta^3(\tilde{k}-\tilde{r})$,
guarantees that the field commutation relation is obeyed -
if $\phi$ obeys
 $\nu(\eta,\tilde{k})\left(
\phi_{\tilde{k}}(\eta) \phi_{\tilde{k}}^{\prime *}(\eta) -
 \phi_{-\tilde{k}}^*(\eta)
\phi_{-\tilde{k}}^\prime(\eta)\right) = i $. 
Normally, this Wronskian condition does not 
determine $\phi$ and therefore $\hat{\phi}$ uniquely, i.e.,   
the choice of a classical
complex solution and correspondingly the choice of
a quantum vacuum are not unique.
Here, however, each $\tilde{k}$ mode
automatically possesses 
a creation time, $\eta_c(\tilde{k})$, at which 
both, $\nu,\mu$ and the 
$\tilde{k}$ mode's equation of motion are singular. This
opens the interesting 
possibility, answering our second question from the beginning,
that the requirement of 
regularity of $\phi$ or other
physical quantities at this singularity 
determines the vacuum uniquely.
\sn
Once the vacuum is fixed, it is then possible to
calculate arbitrary quantum field theoretic entities, 
such as the magnitude of
$\langle 0\vert \hat{\phi}^\dagger_{\tilde{k}}
\hat{\phi}_{\tilde{k}}\vert 0\rangle$ 
after horizon crossing, which
yields the prediction for the
CMB perturbation spectrum. 
This was our third question, and it
can certainly be addressed at least numerically.
The results
should be very interesting to
compare with
the standard inflationary prediction 
of near scale invariance and 
canonically too large amplitude.
\sn
{\bf Acknowledgement}:
The author is happy to thank J. C. Niemeyer, J. R. Klauder,
Ch. Thorn, K. Bering and W. H. Kinney for valuable
 suggestions and comments.

\end{document}